\RequirePackage{ifpdf}
\ifpdf 
\documentclass[pdftex]{sigma}
\else
\documentclass{sigma}
\fi

\newcommand{\bex}{\begin{exa}}
\newcommand{\eex}{\end{exa}}
\newcommand{\br}{\begin{rem}}
\newcommand{\er}{\end{rem}}
\newcommand{\bd}{\begin{Def}}
\newcommand{\ed}{\end{Def}}
\newcommand{\bt}{\begin{theorem}}
\newcommand{\et}{\end{theorem}}
\newcommand{\bl}{\begin{lemma}}
\newcommand{\el}{\end{lemma}}

\newcommand{\adots}{\mathinner{\mkern2mu\raise1pt\hbox{.}\mkern2mu
\raise4pt\hbox{.}\mkern2mu\raise7pt\hbox{.}\mkern1mu}}

\newcommand{\bear}{\begin{array}}
\newcommand{\eear}{\end{array}}
\newcommand\la{{\lambda}}

\newcommand\om{{\omega}}

\newcommand\dd{\mathrm{d}}
\newtheorem{thm}{Theorem}[section]

\newtheorem{rem}[thm]{Remark}

\newtheorem{exa}[thm]{Example}

\newtheorem{lem}[thm]{Lemma}

\theoremstyle{remark}

\newcommand{\Z}{{\mathbb Z}}
\newcommand{\C}{{\mathbb C}}
\newcommand{\R}{{\mathbb R}}

\numberwithin{equation}{section}

\begin{document}

\allowdisplaybreaks

\renewcommand{\thefootnote}{$\star$}

\renewcommand{\PaperNumber}{091}

\FirstPageHeading

\ShortArticleName{Symplectic Maps from Cluster Algebras}

\ArticleName{Symplectic Maps from Cluster Algebras\footnote{This
paper is a contribution to the Proceedings of the Conference ``Symmetries and Integrability of Dif\/ference Equations (SIDE-9)'' (June 14--18, 2010, Varna, Bulgaria). The full collection is available at \href{http://www.emis.de/journals/SIGMA/SIDE-9.html}{http://www.emis.de/journals/SIGMA/SIDE-9.html}}}

\Author{Allan P.~FORDY~$^\dag$ and Andrew HONE~$^\ddag$}

\AuthorNameForHeading{A.P.~Fordy and A.~Hone}

\Address{$^\dag$~School of Mathematics,
University of Leeds, Leeds LS2 9JT, UK}
\EmailD{\href{mailto:a.p.fordy@leeds.ac.uk}{a.p.fordy@leeds.ac.uk}}
\URLaddressD{\url{http://www.maths.leeds.ac.uk/cnls/research/fordy/fordy.html}}

\Address{$^\dag$~School of Mathematics, Statistics and Actuarial Science,
University of Kent,\\
\hphantom{$^\dag$}~Canterbury CT2 7NF, UK}
\EmailD{\href{mailto:A.N.W.Hone@kent.ac.uk}{A.N.W.Hone@kent.ac.uk}}
\URLaddressD{\url{http://www.kent.ac.uk/IMS/staff/anwh/index.html}}

 \ArticleDates{Received May 16, 2011, in f\/inal form September 16, 2011;  Published online September 22, 2011}

 \Abstract{We consider nonlinear recurrences generated from the iteration of maps
that arise from cluster algebras. More precisely, starting from a skew-symmetric
integer matrix, or its corresponding
quiver, one can def\/ine a set of
mutation operations, as well as a set of associated cluster mutations
that are applied to a set of af\/f\/ine coordinates (the cluster variables).
Fordy and Marsh recently provided a complete classif\/ication of all such
quivers that have a certain periodicity property under sequences of mutations.
This periodicity implies that a suitable sequence of cluster mutations
is precisely equivalent to iteration of a nonlinear recurrence relation.
Here we explain brief\/ly how to introduce a %
symplectic structure in this setting, which is preserved by a corresponding
birational map (possibly on a space of lower dimension). We give examples of
both integrable and non-integrable maps that arise from this construction.  We
use algebraic entropy as an approach to classifying integrable cases. The
degrees of the iterates satisfy a tropical version of the map.}

\Keywords{integrable maps; Poisson algebra; Laurent property; cluster
algebra; algebraic entropy; tropical}

\Classification{37K10; 17B63; 53D17; 14T05}

\section{Introduction}

The purpose of this short note is to make a preliminary announcement of our
recent results on recurrence relations which arise in the context of cluster
mutations.  More details and many more examples will be published in a later
paper~\cite{fordy_hone_inprep}.

The theory of cluster algebras was introduced by Fomin and Zelevinsky
\cite{fz1}.  One can start with an $N\times N$, skew-symmetric integer matrix
$B=(b_{jk})\in \mathrm{Mat}_N(\Z )$, which def\/ines a quiver~$Q$ with $N$ nodes
having no 1-cycles or 2-cycles.  This is a directed graph specif\/ied by the rule
that $b_{jk}=-b_{kj}$ denotes the number of arrows from node $j$ to node $k$
(with an overall minus sign corresponding to reversing the direction of the
arrows). Starting from the quiver $Q$ and its associated matrix $B=B(Q)$ one
can def\/ine a set of operations $\mu_k$, called mutations, associated to the
vertices of $Q$ (labelled by $k=1,\ldots ,N$), each of which produces a new
quiver $\tilde{Q}=\mu_kQ$, together with its associated matrix
$\tilde{B}=B(\tilde{Q})$, whose components are given by
\begin{gather*}
\tilde{b}_{j\ell} = \left\{
\begin{array}{ll}
 -b_{j\ell},\qquad & \mathrm{if} \quad
j=k\quad  \mathrm{or}\quad \ell=k, \vspace{1mm}\\
 b_{j\ell} + \frac{1}{2}(|b_{jk}|b_{k\ell} + b_{jk}|b_{k\ell}|), & \mathrm{otherwise.}
\end{array}  \right.
\end{gather*}

In this theory, an af\/f\/ine coordinate $x_j$, $j=1,\ldots ,N$, is associated to
each vertex of the quiver $Q$, and
the set of these $N$ coordinates
def\/ines an initial cluster. Together with
the quiver mutations (and the above equivalent set of matrix mutations), there
is an associated set of cluster mutations, which replace the initial cluster
$(x_1,\ldots , x_N)$ with a new cluster $(\tilde{x}_1,\ldots , \tilde{x}_N)$,
where the new variables are related to the old ones by a simple birational
transformation:
\begin{gather*} 
\tilde{x}_k x_k = \prod_{j=1}^N x_j^{[b_{k,j}]_+} + \prod_{j=1}^N
x_j^{[-b_{k,j}]_+}, \qquad \tilde{x}_j= x_j, \qquad j\neq k,
\end{gather*}
where the square bracket notation is def\/ined by $[b]_+=\max (b,0)$. This map is
an involution, because applying the mutation $\mu_k$ once more restores the
original cluster ${\bf x}=(x_j)$ and also sends~$\tilde{B}$ back to~$B$.
(Note that in this paper we do not consider the most general version of
cluster algebras, but only coef\/f\/icient-free algebras for which the
exchange matrix~$B$ is skew-symmetric.)

The full set of cluster variables, obtained by all possible sequences of
mutations, generates the cluster algebra. However, in certain cases, which were
classif\/ied recently by Fordy and Marsh~\cite{fordy_marsh}, the quiver has a
special periodicity property, which means that a particular sequence of
mutations is equivalent to generating the cluster variables via iteration of a
single recurrence relation of the form
\begin{gather}\label{crec}
x_{n+N}x_n =\prod_{j=1}^{N-1} x_{n+j}^{[b_{1,j+1}]_+}+\prod_{j=1}^{N-1}
x_{n+j}^{[-b_{1,j+1}]_+},
\end{gather}
where the exponents $\pm b_{1,k}$ are the elements of the top row of the
skew-symmetric integer mat\-rix~$B$. The results of~\cite{fordy_marsh}
imply that in order for this periodicity property to hold,
the vector $(b_{12},\ldots ,b_{1N})$
of exponents in the top row must be palindromic, but is otherwise
arbitrary, and the rest of matrix $B$
is completely determined from the these exponents.

By a theorem of Fomin and Zelevinsky~\cite{fz1},
recurrences which are constructed in this way are guaranteed to have the
Laurent property, meaning that the iterates are Laurent polynomials
in the initial data $x_1,\ldots ,x_N$ with integer coef\/f\/icients
(but see~\cite{fz2} for a more general treatment of the Laurent phenomenon).
One of the earliest examples of a rational (nonlinear) recurrence having
the Laurent property is known as Somos-5.

\bex[The Somos-5 recurrence] \rm
The Somos-5 recurrence is
\begin{gather} \label{somos5}
x_{n+5}  x_n = x_{n+4}  x_{n+1}+ x_{n+3}  x_{n+2}.
\end{gather}
As shown in \cite{fordy_marsh}, this recurrence corresponds to the cluster
exchange relation generated by the node labelled by $j=1$ in a quiver $Q$ with
f\/ive nodes, that is equivalent to the skew-symmetric $5\times 5$ integer matrix
\begin{gather} \label{s5bmatrix} %
B=\begin{pmatrix}
0 & -1 & 1 & 1 & -1 \\
1 & 0 & -2 & 0 &  1 \\
-1 & 2 & 0 & -2 & 1 \\
-1 & 0 & 2 & 0 & -1 \\
1 & -1 & -1 & 1 & 0 \end{pmatrix}.
\end{gather}
The mutation $\mu_1$ transforms this
matrix to
\[
\tilde{B} = \mu_1 B =  \begin{pmatrix}
0 & 1 & -1 & -1 & 1 \\
-1 & 0 & -1 & 1 &  1 \\
1 & 1 & 0 & -2 & 0 \\
1 & -1 & 2 & 0 & -2 \\
-1 & -1 & 0 & 2 & 0 \end{pmatrix},
\]
which, in this case, turns out to be the permutation of indices
$(1,2,3,4,5)\mapsto (5,1,2,3,4)$. In other words, $\tilde{B}=\rho B\rho^{-1}$,
where
\[
\rho = \begin{pmatrix}
0 & 0 & 0 & 0 & 1 \\
1 & 0 & 0 & 0 &  0 \\
0 & 1 & 0 & 0 & 0 \\
0 & 0 & 1 & 0 & 0 \\
0 & 0 & 0 & 1 & 0 \end{pmatrix}
\]
is the matrix representing this cyclic permutation. At the same time, the
associated transformation of the cluster variables is
$(x_1,x_2,x_3,x_4,x_5)\mapsto (\tilde{x}_1,x_2,x_3,x_4,x_5)$, where
$\tilde{x}_1$ is def\/ined by the exchange relation
\[
\tilde{x}_1x_1 = x_2x_5 + x_3x_4.
\]
Because the f\/irst matrix mutation $\mu_1$ just corresponds to a cyclic
permutation of the indices, it follows that a subsequent mutation $\mu_2$ of
the new matrix $\tilde{B}$ produces the same formula for the exchange relation
(up to relabelling the variables). It is this property which allows an inf\/inite
sequence of mutations to be considered as equivalent to the iteration of a
single recurrence relation (in this case, the recurrence~(\ref{somos5}) above).
 \eex

\br[Laurent phenomenon]  \rm
The Somos-5 recurrence is distinguished for several reasons. It is one of the
f\/irst known examples of a rational recurrence whose iterates are Laurent
polynomials in the initial values with integer coef\/f\/icients, which for
(\ref{somos5}) means that
\[
x_n \in \Z \big[x_1^{\pm 1},  x_2^{\pm 1},x_3^{\pm 1},x_4^{\pm 1},x_5^{\pm 1}\big]
\]
for all $n$. This Laurent phenomenon was noted by Michael Somos, who
saw that
it explained his earlier observation that the recurrence (\ref{somos5})
generates an integer sequence starting from the initial data
$(x_1,\dots,x_5)=(1,1,1,1,1)$, that is
\[
1,1,1,1,1,2,3,5,11,37,83,274,1217,6161,22833,\ldots .
\]
The latter sequence also appears in a problem of Diophantine geometry, and
Elkies and others explained the connection with elliptic curves and theta
functions (see~\cite{heron}). Excellent accounts of the earliest results on the
Laurent property appear in the articles of Gale~\cite{gale}. However, a more
complete picture did not begin to emerge until the work of Fomin and
Zelevinksy, who were able to treat a wide variety of examples of the Laurent
phenomenon on the same footing~\cite{fz2}.
\er

Another reason to highlight the Somos-5 recurrence is the fact that it is
connected to one of the archetypal examples of a discrete integrable system,
namely a member of the Quispel--Roberts--Thompson (QRT) family of maps of the
plane~\cite{qrt}.  This is described in Example~\ref{somos5pbex} below.

The main motivation behind \cite{fordy_hone_inprep} (and the current
announcement) is to explain the connection between discrete integrable systems
and cluster algebras, and try to understand to what extent examples like
Somos-5 are isolated rarities. However, in order to be able to talk about
Liouville--Arnold integrability in the f\/inite-dimensional setting, it is
necessary to have a symplectic (or Poisson) structure to hand. In the next
section we describe how this arises in the context of cluster algebras, and use
Somos-5 for illustration.

A further example that reduces to a map of the plane is introduced and evidence
given for its non-integrability. In the third section we consider recurrences
of the form (\ref{crec}) from the viewpoint of their algebraic entropy  (as
def\/ined in \cite{bellon_viallet}), and suggest that this leads to a very sharp
classif\/ication result for these systems.

The f\/inal section is reserved for some conclusions.

\section{Poisson and symplectic structures}

Cluster transformations were considered in the setting of Poisson geometry by
Gekhtman, Shapiro and Vainshtein \cite{gsv}, who found Poisson structures of
log-canonical type for the cluster variables:
\begin{gather}\label{logcan} %
\{x_j,x_k\}=c_{jk}\, x_{j}x_k,
\end{gather}
for some constant skew-symmetric matrix $C=(c_{jk})$. Such Poisson brackets are
compatible with cluster mutations, in the sense that the new cluster obtained
by mutation also satisf\/ies a~log-canonical bracket:
\[
\{\tilde{x}_j,\tilde{x}_k\}=\tilde{c}_{jk}\, \tilde{x}_{j}\tilde{x}_k.
\]
However, in general $\tilde{C}\neq C$, so the mutation map
from the variables ${\bf x}$ to $\tilde{{\bf x}}$ (which can be considered as
an involution of $N$-dimensional af\/f\/ine space)
does not preserve the original Poisson structure.

Nevertheless, the recurrences def\/ined by the special sequence of mutations
available in the case of mutation-periodic exchange matrices {\em do}
correspond to maps with an invariant Poisson bracket of log-canonical form
(perhaps on a lower-dimensional symplectic manifold). Theorem~\ref{torusred} below describes how this works.

Given the recurrence (\ref{crec}), we def\/ine a birational map   $\varphi:
\C^N\rightarrow \C^N$:
\begin{gather} \label{bir} %
\varphi: \  \begin{pmatrix}
x_1 \\
x_2 \\
\vdots  \\
x_{N-1} \\
x_{N}
\end{pmatrix}
\longmapsto \begin{pmatrix}
x_2 \\
x_3 \\
\vdots  \\
x_{N} \\
x_{N+1}
\end{pmatrix},
\qquad \mathrm{where} \quad x_{N+1}=\frac{ \prod\limits_{j=1}^{N-1} x_{j+1}^{[b_{1,j+1}]_+} +
\prod\limits_{j=1}^{N-1} x_{j+1}^{[-b_{1,j+1}]_+}
 }{x_1}.
\end{gather}
It is possible to seek a skew-symmetric matrix $C=(c_{jk})$ such that the
Poisson bracket (\ref{logcan}) is invariant.
The nature of the map (being of
the form $\tilde x_j=x_{j+1}$)
means that this matrix has a~banded structure,
so only the top row needs to be determined.

\bex[Invariant Poisson bracket for the Somos-5 map] \label{somos5pbex} \rm
The Somos-5 map is
 \begin{gather} \label{birs5}
\varphi: \
\begin{pmatrix}
x_1 \\
x_2 \\
x_3 \\
x_4  \\
x_{5}
\end{pmatrix}
\longmapsto \begin{pmatrix}
x_2 \\
x_3 \\
x_4  \\
x_{5} \\
\displaystyle \frac{x_2x_5+x_3x_4}{x_1}
\end{pmatrix},
\end{gather} which preserves the log-canonical Poisson bracket
\begin{gather} \label{logcans5}  %
\{x_j,x_k\}=(j-k)  x_{j}x_k,
\end{gather}
in the sense that $\{\varphi^* F, \varphi^* G\} = \varphi^* \{ F,G \}$ for any
pair of functions~$F$,~$G$. The corresponding matrix
\[
C=\begin{pmatrix}
0 & -1 & -2 & -3 & -4 \\
1 & 0 & -1 & -2 & -3 \\
2 & 1 & 0 & -1 & -2 \\
3 & 2 & 1 & 0 & -1 \\
4 & 3 & 2 & 1 & 0
\end{pmatrix}
\]
has rank 2, and the three independent null vectors
\[
{\bf m}_1=(1,-2,1,0,0)^T, \qquad
{\bf m}_2=(0,1,-2,1,0)^T, \qquad
{\bf m}_3=(0,0,1,-2,1)^T
\]
provide three independent Casimir functions for the bracket, $f_j = {\bf
x}^{{\bf m}_j}$ for $j=1,2,3$:
\[
f_1 = \frac{x_1 x_3}{x_2^2}, \qquad
f_2 = \frac{x_2 x_4}{x_3^2}, \qquad
f_3 = \frac{x_3 x_5}{x_4^2}.
\]
However, under the action of $\varphi$, as in (\ref{birs5}), the Casimirs transform as
\[
\varphi^* f_1 = f_2,\qquad \varphi^* f_2 = f_3, \qquad \varphi^* f_3 =
\frac{f_2f_3+1}{f_1f_2^2f_3^2} .
\]
This induced action of $\varphi$ on the variables $f_j$ was
written in terms of a third-order recurrence in~\cite{hones5}.
\br[Non-invariance of symplectic leaves]  \rm
We see that $\varphi$ does not preserve the symplectic leaves of the bracket
because the Casimirs are not invariant under the map.
It follows that the
log-canonical bracket on the cluster variables $x_j$ is not relevant to the
integrability of the map (\ref{somos5}).
\er

In fact, the variables
\begin{gather} \label{yvars} %
y_1=\frac{x_1x_4}{x_2x_3}, \qquad y_2 = \frac{x_2x_5}{x_3x_4}
\end{gather}
which are themselves Casimirs of this bracket, since $y_1=f_1f_2$ and
$y_2=f_2f_3$, transform as a~map of the plane, given by
\begin{gather}\label{qrtmap}  %
\hat{\varphi}: \  \begin{pmatrix} y_1 \\ y_2 \end{pmatrix} \mapsto
\begin{pmatrix} y_2 \\ (y_2+1)/(y_1y_2) \end{pmatrix}.
\end{gather}
This is a particular case of the QRT map \cite{qrt} and possesses an invariant
Poisson bracket of log-canonical type:
\begin{gather}\label{logcanqrt}  %
\{y_1,y_2\}=y_1y_2.
\end{gather}
Moreover, the quantity
\[
H=y_1+y_2 + \frac{1}{y_1} + \frac{1}{y_2}+ \frac{1}{y_1y_2}
\]
is an invariant of the map, which means that~(\ref{qrtmap}) is integrable in
the Liouville--Arnold sense~\cite{veselov}. The smooth level curves of $H$ have
genus one, which explains the connection with elliptic curves and leads to the
explicit solution of the initial value problem for~(\ref{somos5}) in terms of
the Weierstrass sigma function~\cite{hones5}.
 \eex

\subsection[Symplectic form from the $B$ matrix]{Symplectic form from the $\boldsymbol{B}$ matrix}

The above construction of the Poisson brackets (\ref{logcans5}) and
(\ref{logcanqrt}) was presented in an ad hoc way.  The matrix
$C=(c_{jk})$ for the map of the coordinates $x_j$, as well as the
bracket between the $y_j$, can be derived purely from the assumption
that a log-canonical bracket exists for each of the maps ${\varphi}$
and $\hat{\varphi}$.
Furthermore, these calculations start from the {\em
map} and have {\em no connection} with the cluster construction.  We now
explain how to derive the relevant Poisson brackets from the exchange matrix
$B$.

We def\/ine the log-canonical two-form associated to $B$, by
\begin{gather} \label{omega}  %
\om =\sum_{j<k} \frac{b_{jk}}{x_jx_k}\dd x_j\wedge \dd x_k ;
\end{gather}
this two-form was f\/irst introduced in \cite{gsvduke}, and was also considered
in \cite{fockgon}.
Let $\varphi$ be the map~(\ref{bir}), associated with the matrix mutation
$\mu_1$ of $B$.  Then we have the following:
\begin{lem} \label{symplectic}
Let $B$ be a skew-symmetric integer matrix. The following conditions are
equivalent.
\begin{enumerate}\itemsep=0pt
\item The matrix $B$ def\/ines a cluster mutation-periodic quiver with period~$1$.
\item  The matrix elements $B$ satisfy the relations
\begin{gather}\label{reln1} %
b_{j,N}=b_{1,j+1}, \qquad j=1,\ldots , N-1,
\end{gather}
and
\begin{gather}\label{reln2} %
b_{j+1,k+1}=b_{j,k}+b_{1,j+1} [-b_{1,k+1}]_+  - b_{1,k+1}
[-b_{1,j+1}]_+  ,
\end{gather} %
for $1\leq j,k \leq N-1$.  %
\item The two-form $\om$ is preserved by the map $\varphi$, i.e. $\varphi^* \om =\om$.
\end{enumerate}
\end{lem}

The above result follows from a direct calculation, which will be
presented in \cite{fordy_hone_inprep}; it can also be understood
in the light of Theorem~2.1 in \cite{gsvduke}, which shows
that the two-form (\ref{omega})
is {\em covariant} with respect to general cluster transformations. We emphasise
that in the case of {\em period $1$ quivers}, this two-form is {\em invariant}.

Note that the relations (\ref{reln1}) and (\ref{reln2}) mean that the vector
$(b_{12},\ldots , b_{1N})=(a_1,\ldots , a_{N-1})$ must be palindromic,
i.e.~$a_j = a_{N-j}$ for all $j$, and
the matrix $B$ can be completely reconstructed from these exponents
in the f\/irst row (see Theorem~6.1 of~\cite{fordy_marsh}).

The case where $B$ is degenerate is more delicate, but the general situation is
described by the following.
\begin{thm} \label{torusred}
The map $\varphi$ is symplectic whenever $B$ is nondegenerate. In the
degenerate case that $\mathrm{rank} \, B = 2K<N$, there is a rational map $\pi$
and a symplectic birational map
$\hat\varphi$ with symplectic form $\hat\om$ such that the 
diagram
\begin{gather*} 
\begin{CD}
\C^N @>\varphi >> \C^N\\
@VV\pi V @VV\pi V\\
\C^{2K} @>\hat{\varphi}>> \C^{2K}
\end{CD}
\end{gather*}
is commutative, and the symplectic form $\hat\om$ on $\C^{2K}$ satisfies
$\pi^*\hat\om = \om$.
\end{thm}

The latter result can be viewed as a special case of the
symplectic reduction of the
form (\ref{omega}) presented in \cite{gsvduke}.

\bex[Somos-5 map]  \label{somos5mapex}  \rm
We consider the 2-form (\ref{omega}) with matrix $B$ given by
(\ref{s5bmatrix}).  Written in terms of the variables $z_i=\log x_i$, we have
\[
\omega = \sum_{j<k} b_{jk}dz_j\wedge dz_k.
\]
The rank of matrix $B$ is 2, since it has three independent null vectors
\[
{\bf u}_1 =(1,1,1,1,1)^T, \qquad {\bf u}_2 =(1,2,3,4,5)^T, \qquad {\bf u}_3
=(1,-1,1,-1,1)^T.
\]
$\mbox{Im}\, B$ is spanned by
\[
{\bf v}_1 =(1,-1,-1,1,0)^T, \qquad {\bf v}_2 =(0,1,-1,-1,1)^T,
\]
and ${\bf u}_i\cdot{\bf v}_j=0$, for all $i$, $j$. This means that
\[
\omega = (dz_2-dz_3-dz_4+dz_5)\wedge (dz_1-dz_2-dz_3+dz_4),
\]
which, in $x$-coordinates, gives
\[
\omega = d \log\left(\frac{x_2x_5}{x_3x_4}\right)\wedge d
\log\left(\frac{x_1x_4}{x_2x_3}\right)=\frac{dy_2\wedge dy_1}{y_1y_2},
\qquad\mbox{where}\quad y_1=\frac{x_1x_4}{x_2x_3},\quad
y_2=\frac{x_2x_5}{x_3x_4}.
\]
We see that on the two-dimensional space with coordinates $y_1$, $y_2$, this
gives a form $\hat\omega$ which is non-degenerate, so def\/ines an invariant
symplectic form.

We are thus led directly to the variables (\ref{yvars}), which satisfy the QRT
map (\ref{qrtmap}), and to the Poisson bracket (\ref{logcanqrt}), which is
def\/ined through the inverse of the symplectic form.
In the general case, the analogues of the variables (\ref{yvars})
correspond to particular choices of the so-called
$\tau$-coordinates introduced in \cite{gsv}.
 \eex

\bex[A sixth-order recurrence]\label{order6ex}  \rm
The recurrence
\begin{gather} \label{sixthordernew} %
x_{n+6}  x_n = x_{n+5}^2 x_{n+3}^4 x_{n+1}^2+(x_{n+4}  x_{n+2})^6,
\end{gather} %
comes from the degenerate exchange matrix
\begin{gather}\label{6b}
B=\begin{pmatrix} 0 & -2 & 6 & -4 & 6 & -2 \\
                      2 & 0 &  -14 & 6 & -16 & 6 \\
                     -6 & 14 & 0  &  10 & 6 & -4 \\
                      4 & -6 & -10 & 0 & -14 & 6 \\
                      -6 & 16 & -6 & 14 & 0 & -2 \\
                       2 & -6 & 4 & -6 & 2 & 0 \end{pmatrix}.
\end{gather}  %
Clearly, iteration of this recurrence is equivalent to iterating the map
$\varphi:  \C^6 \to \C^6$ given by
\[
\varphi: \  (x_1,x_2,x_3,x_4,x_5,x_6)\mapsto (x_2,x_3,x_4,x_5,x_6,x_7), \qquad
x_7 =\frac{x_6^2x_4^4x_2^2 + x_5^6x_3^6}{x_1},
\]
and this map preserves the degenerate two-form given by the expression
(\ref{omega}) with coef\/f\/icients~$b_{jk}$ as in~(\ref{6b}).

Written in terms of the variables $z_i=\log x_i$, we have
\[
\omega = \sum_{j<k} b_{jk}dz_j\wedge dz_k.
\]
The rank of matrix $B$ is 2, since it has four independent null vectors
\begin{alignat*}{3}
& {\bf u}_1 =(1,0,-1,0,1,0)^T, \qquad && {\bf u}_2 =(0,1,0,-1,0,1)^T,& \\
& {\bf u}_3
=(3,1,0,0,0,-1)^T, \qquad && {\bf u}_3 =(-1,0,0,0,1,3)^T.&
\end{alignat*}
$\mbox{Im}\, B$ is spanned by
\[
{\bf v}_1 =(1,-3,2-3,1,0)^T, \qquad {\bf v}_2 =(0,1,-3,2-3,1)^T,
\]
and ${\bf u}_i\cdot{\bf v}_j=0$, for all $i$, $j$. This means that
\[
\omega = 2(dz_2-3dz_3+2dz_4-3dz_5+dz_6)\wedge (dz_1-3dz_2+2dz_3-3dz_4+dz_5),
\]
which, in $x$-coordinates, gives
\begin{gather*}
\omega =2 d \log\left(\frac{x_2x_4^2x_6}{x_3^3x_5^3}\right)\wedge d
\log\left(\frac{x_1x_3^2x_5}{x_2^3x_4^3}\right)=2\frac{dy_2\wedge
dy_1}{y_1y_2},
\end{gather*}
where
\begin{gather*}
y_1=\frac{x_1x_3^2x_5}{x_2^3x_4^3},\qquad
y_2=\frac{x_2x_4^2x_6}{x_3^3x_5^3}.
\end{gather*}
We see that on the 2 dimensional space with coordinates $y_1$, $y_2$, $\omega$ is
non-degenerate, so def\/ines a symplectic form, which is invariant under the
induced map
\begin{gather}\label{sixthmap}
\hat{\varphi}: \  \begin{pmatrix} y_1 \\ y_2 \end{pmatrix} \mapsto
\begin{pmatrix} y_2 \\ (y_2^2+1)/(y_1y_2^3) \end{pmatrix} .
\end{gather}
However, in this case we know of no f\/irst integral and, furthermore, numerical
plots indicate the presence of chaos.
 \eex

In the next section we discuss algebraic entropy and show that the algebraic
entropy of this particular map is non-zero.

\section{Algebraic entropy}

Bellon and Viallet \cite{bellon_viallet} proposed a suitable measure of entropy
for rational maps.  This counts the degree $d_n$ of the $n$th iterate of such a
map in dimension $N$, def\/ined as the maximum degree of the iterates,
considered as rational functions of the $N$ coordinates corresponding to the
initial data.  The algebraic entropy $\mathcal{E}$ of the map is def\/ined as
\[
\mathcal{E}=\lim_{n\to\infty}\frac{1}{n}\log d_n.
\]
For a generic map of degree $d$, the entropy is $\log d>0$, but for special
maps cancellations of factors from the numerators and denominators of the
rational functions can occur upon iteration, so that the entropy is smaller
than expected.

It has been observed (see \cite{bellon_viallet} for a discussion) that rational
maps which are Liouville--Arnold integrable have zero algebraic entropy and it
is
{\em conjectured} that zero algebraic entropy implies integrability. In any
case, it appears to be a reliable indicator of integrability.
In the setting of this paper, Liouville--Arnold integrability
would require the existence of $K$ independent functions
of the variables $y_1,\ldots, y_{2K}$, which are invariant under the action
of the map $\hat\varphi$ and in involution with respect to the
Poisson bracket def\/ined by $\hat\om$.

In general it can be dif\/f\/icult to calculate the entropy of a rational map
explicitly. Even numerical calculations become dif\/f\/icult if the dimension is
large, since it is necessary to perform a symbolic computation of the rational
functions generated by the map and count their degrees. However, for the family
of maps of the form (\ref{bir}) (or equivalently the recurrences (\ref{crec})),
there is a considerable simplif\/ication.
From the Laurent property of the associated cluster algebra~\cite{fz1}, it
follows that the iterates $x_n$ are Laurent polynomials in the initial
conditions $x_1, \ldots , x_N$.  Thus it is suf\/f\/icient to count the degrees of
the monomials that appear as the denominators of the iterates.

Upon writing each iterate in lowest terms as $x_n = D_n({\bf x})/M_n({\bf x})$
(where, as before, the vector ${\bf x}=(x_j)$ denotes the variables of the
initial cluster, and  $x_j \nmid D_n({\bf x})$ for any $j=1,\ldots ,N$), the
denominator is a monomial $M_n = {\bf x}^{{\bf d}_n}$, with ${\bf
d}_n=\big(d_n^{(1)},\ldots ,d_n^{(N)}\big)^T$ being the vector of degrees. Substituting
this form of the Laurent polynomials into (\ref{crec}) and comparing
denominators on each side, it is apparent that (for suf\/f\/iciently large $n$) the
degrees $d_n^{(j)}$ satisfy the same recurrence for all $j$, namely
\begin{gather} \label{tropical} %
d_{n+N}+d_n = \max \left(   \sum_{j=2}^N [b_{1,j}]_+
d_{n+j-1}  , \, \sum_{j=2}^N  [-b_{1,j}]_+  d_{n+j-1} \right) .
\end{gather} %
The above equation is precisely the ultra-discrete version of (\ref{crec}) (an
example of tropical ma\-the\-matics \cite{sturmfels}, where the ordinary f\/ield
operations are replaced by the max-plus algebra). The entropy of these maps is
measured by the growth of the degrees of the denominators, and generically the
total degree, $\mathrm{deg} \, M_n=\sum\limits_{j=1}^Nd_n^{(j)}$, grows exponentially
at the same rate as each component $d_n^{(j)}$, which is controlled by the
tropical version of the map, as in (\ref{tropical}).

\bex[Tropical Somos-5] \rm
For the Somos-5 recurrence, the ultra-discrete analogue of the bilinear
equation  (\ref{somos5}) is
\begin{gather}\label{ts5}
d_{n+5}+ d_n = \max \left(
d_{n+4}+d_{n+1}, d_{n+3}+d_{n+2}\right).
\end{gather} %
The singularity analysis of ultra-discrete maps has been considered quite
recently \cite{ultra}. For this particular example, it is possible to describe
the general solution quite explicitly. Indeed, the tropical analogue of the
quantity $y_n$ (for the recurrence version of the QRT map (\ref{qrtmap})) is
\[
Y_n=d_{n+3}+d_n-d_{n+2}-d_{n+1},
\]
which satisf\/ies the second order relation
\begin{gather}\label{tqrt}
Y_{n+2}+Y_n=[Y_{n+1}]_+ - Y_{n+1}.
\end{gather}  %
The latter is the tropical version of the QRT map def\/ined in (\ref{qrtmap}).
Although in the context described above $d_n$ counts the degree of a monomial,
so that only non-negative integer values of $d_n$ arise, which lead to  $Y_n$
taking integer values only, the recurrence (\ref{tqrt}) can be considered more
generally as def\/ining a piecewise linear map  $\R^2\to\R^2$. For this map it
can be verif\/ied directly that all the orbits are periodic with period 7, from
which it follows that $d_n$ satisf\/ies a~linear recurrence of order 10 with
constant coef\/f\/icients, namely
\begin{gather}\label{ts5lin} %
\big(\mathcal{S}^7-1\big)( d_{n+3}+d_n-d_{n+2}-d_{n+1})=0,
\end{gather}  %
where we have used $\mathcal{S}$ to denote the shift operator, def\/ined by
$\mathcal{S}  f_n =f_{n+1}$ for any function $f$ of $n\in\Z$. (Note that
the periodicity of orbits in ultradiscrete QRT maps was proved by Nobe in~\cite{nobe}.)

For the sequence of degrees of the tropical Somos-5 recurrence (\ref{ts5}),
in each variable $x_j$ the denominator has four steps of degree zero, say $d_1=d_2=d_3=d_4=0$,
before the f\/irst non-zero degree $d_5=1$, and these f\/ive initial conditions generate
a sequence beginning
\[
0,0,0,0,1,1,1,2,3,3,4,5,6,6,8,9,10,11,13,14,15,17,19,20,22,24,26,27,30,\ldots ,
\]
which has quadratic growth with $n$, indicating that the algebraic entropy is zero.
To see that $d_n=O(n^2)$, it is enough to note that the above sequence satisf\/ies the tenth
order linear recurrence~(\ref{ts5lin}), whose characteristic polynomial
is $(\la^7-1)(\la ^3 -\la^2 -\la +1)$, and all of the roots of the latter have
modulus 1, with $\la =1$ being a triple root.
 \eex  %

\bex[Tropical sixth-order recurrence] \label{order6trpex}  \rm
The example of the recurrence (\ref{sixthordernew}) is quite dif\/ferent, in that
the sequence of degrees, which begins
\[
0,0,0,0,0,1,2,6,16,42,110,287,754,1974,5168,13530,35422,92737,242788,635628,
\ldots ,
\]
grows exponentially with $n$. In this case, the degrees satisfy the tropical
recurrence
\begin{gather*}
d_{n+6}+ d_n = \max \left( 2d_{n+5}+4d_{n+3}+2d_{n+1},
6(d_{n+4}+d_{n+2})\right),
\end{gather*} %
while the analogue of  (\ref{sixthmap}) is the piecewise linear map on $\R^2$ given by
\begin{gather} \label{tsixthmap} %
\begin{pmatrix}
Y_1 \\
Y_2
\end{pmatrix}
\longmapsto \begin{pmatrix}
Y_2 \\
2\left[ Y_2 \right]_+
-3Y_2-Y_1
\end{pmatrix},
\end{gather}
where
\[
Y_n = d_{n+4}+2d_{n+2}+d_n - 3d_{n+3}-3d_{n+1}.
\]
The particular orbit of the map for $Y_n$ corresponding
to the degree sequence has initial data $(Y_1,Y_2)=(0,1)$,
and the sequence of $Y_n$
begins $0,1,-1,2,-1,1,0,-1,3,-2,3,-1$; thereafter it
is periodic with period 12,
which implies that the degrees $d_n$ satisfy a linear recurrence
of order~16 with constant coef\/f\/icients, that is
\[
\big(\mathcal{S}^{12} -1 \big)( d_{n+4}+2d_{n+2}+d_n - 3d_{n+3}-3d_{n+1})=0.
\]
(However, note that the general orbit of the map
(\ref{tsixthmap}) is not periodic.)
The right hand bracket above contributes a factor
of $\la^4-3\la^3+2\la^2 -3\la +1 = (\la^2+1)(\la^2 -3\la +1)$ to the
characteristic polynomial, and this contains the root of
largest magnitude, namely $\la_{\max} = (3+\sqrt{5})/2$.
This implies that $\log d_n\sim n \log \la_{\max}$
as $n\to\infty$, giving a positive value to the algebraic entropy,
$\mathcal{E}=  \log \la_{\max}$.
 \eex

Although we are not aware of any theorems that directly connect algebraic
entropy to the Liouville--Arnold notion of integrability for maps, the fact that
a map has $\mathcal{E}>0$ is a strong indicator of non-integrability. For each
of the recurrences (\ref{crec}), the degrees of each variable appearing in the
denominators of the sequence of Laurent polynomials satisfy the same tropical
recurrence (\ref{tropical}), which means that any set of $N+1$ adjacent degrees
always satisf\/ies one of the two linear relations
\[
d_{n+N}+d_n = \sum_{j=2}^N [\pm b_{1,j}]_+  d_{n+j-1}.
\]
Empirically we have observed that for large enough $n$ the degrees $d_n$
actually switch periodically between the latter two linear recurrences. This
suggests that the growth of the degrees is controlled by whichever of the two
characteristic polynomials
\[
P_{\pm}(\la ) =\la^N + 1 - \sum_{j=1}^{N-1}[ \pm b_{1,j+1}]_+\la^j
\]
has the root $\la_{\max}$ of largest magnitude. It is fairly easy to show that
$|\la_{\max}|>1$  provided that
\begin{gather} \label{maxe}
\max \left(   \sum_{j=2}^N [b_{1,j}]_+   , \, \sum_{j=2}^N
[-b_{1,j}]_+   \right) \geq 3,
\end{gather}
and in that case the entropy should be $\mathcal{E}=  \log \la_{\max}>0$. Thus
we are led to the

\medskip\noindent {\bf Conjecture.} {\it The condition \eqref{maxe}  is sufficient for
positive algebraic entropy.}

\medskip More detailed evidence for this conjecture will be
described in \cite{fordy_hone_inprep}.

\section{Conclusions}

In this paper we have brief\/ly discussed the recurrences (\ref{crec}) (and
corresponding maps (\ref{bir})), which arise naturally in the context of
cluster mutations of mutation-periodic quivers \cite{fordy_marsh}.  Our main
interest is to understand the integrability or otherwise of the resulting maps.
To this end, we have shown that the cluster exchange matrix $B$ def\/ines a
2-form (\ref{omega}), which is invariant under the action of the corresponding
map.  Furthermore, since, for an even number of nodes, the matrix $B$
(corresponding to a generic, period 1 quiver) is non-singular, this 2-form is
generically symplectic, so the inverse matrix def\/ines a non-degenerate,
invariant Poisson bracket of log-canonical type.  However, for any quiver with
an odd number of nodes and for some special (but important) examples with an
even number, the 2-form (\ref{omega}) is degenerate.  However, as we
have seen in the two examples presented here, it is possible to reduce the map
to a lower-dimensional manifold with an invariant symplectic form; this is the
Weil--Petersson form introduced in \cite{gsvduke}.  Again, we emphasise that in
our context, this symplectic form is {\em invariant}, not just {\em covariant}.

Finding the invariant symplectic form (and hence the Poisson bracket) is only
part of the task.  For $K$ degrees of freedom we need $K$ invariant functions
which are in involution with one another.
When $K=1$ it is enough to f\/ind a single invariant
function (as with the QRT map (\ref{qrtmap})).  For $K>1$ it can be much more
dif\/f\/icult.  The simplest period 1 quivers are the primitives (see
\cite{fordy_marsh} for the def\/inition), which are linearisable.  In this case
it is possible \cite{fordy_rkb} to construct a bi-Hamiltonian ladder to prove
complete integrability.  In \cite{fordy_hone_inprep} we present other
linearisable examples which can also be shown to be completely integrable,
although this requires a dif\/ferent construction.

We have used algebraic entropy arguments to show that complete integrability
will be rare for the maps obtained through cluster mutation.
A non-integrable example is provided by the map
(\ref{sixthordernew}), which was reduced to a symplectic map
with one degree of freedom, but this was shown to have positive entropy, and
we have numerical results indicating chaotic behaviour.

The conjectured criterion (\ref{maxe}) for the maps to have
positive algebraic entropy leads to a~sharp classif\/ication of the integrable
maps that can arise from recurrences of the form (\ref{crec}).
From this criterion it follows that systems with vanishing entropy must satisfy
$S_{\pm}\leq 2$, where $S_{\pm}=\sum\limits_{j=2}^N [\pm b_{1,j}]_+$ are the sums of
the positive and negative entries, respectively. By the symmetry $B\to -B$,
which does not change the recurrence, one can take $S_{+}\geq S_{-}$ without
loss of generality. It then follows that the sums of the positive and negative
entries are restricted to take the values
\[
(S_+,S_-)  =   (1,0) , \, (2,0)   , \,(2,1)   , (2,2)
\]
only. Each of these four cases can be subjected to further analysis, to verify
that they correspond to symplectic maps (perhaps on a reduced manifold) that
are integrable in the Liouville--Arnold sense (see \cite{fordy_hone_inprep} for
further details).

\subsection*{Acknowledgments} The authors would like to thank the Isaac
Newton Institute, Cambridge for hospitality during the Programme on Discrete
Integrable Systems, where this collaboration began. They are also grateful to
the organisers of SIDE 9 in Varna for inviting us both to speak there.

\pdfbookmark[1]{References}{ref}
\LastPageEnding

\end{document}